\documentclass[]{aa}
\usepackage{graphicx,natbib}

\def\ms{\mbox{$\rm M_\odot$}}

\begin{document}

\title{The lenticular galaxy NGC\,3607: stellar population, metallicity and ionised gas}

\author{M. G. Rickes\inst{1}, M. G. Pastoriza\inst{1} \and C. Bonatto\inst{1} }

\offprints{C. Bonatto - charles@if.ufrgs.br}

\institute{Universidade Federal do Rio Grande do Sul (UFRGS), Instituto de F\'\i sica,
 Departamento de Astronomia\\
\email{maurogr@if.ufrgs.br; mgp@if.ufrgs.br and charles@if.ufrgs.br} }

\date{Received --; accepted --}

\abstract
{The investigation of the dominant formation mechanism of early-type galaxies.}
{In this work we derive clues to the formation scenario and ionisation source
of the lenticular galaxy NGC\,3607 by means of metallicity gradients, stellar 
population and emission lines properties.}
{We work with long-slit spectroscopy from which we (i) study the radial distribution
of the equivalent widths of conspicuous metallic absorption features, (ii) infer on
the star-formation history (with a stellar population synthesis algorithm),
and (iii) investigate the ionisation source responsible for a few strong emission
lines.}
{Negative radial gradients are observed for most of the absorption features of NGC\,3607.
Compared to the external parts, the central region has a deficiency of alpha elements and 
higher metallicity, which implies different star-formation histories in both regions. At
least three star formation episodes are detected, with ages within 1-13\,Gyr. The dynamical 
mass and the $Mg_2$ gradient slope are consistent with mergers being important contributors
to the formation mechanism of NGC\,3607. Emission-line ratios indicate the presence of a 
LINER at the centre of NGC\,3607. Contribution of hot, old stars to the gas ionisation 
outside the central region is detected.}
{Evidence drawn from this work suggest small mergers as important contributors to
the formation of NGC\,3607, a scenario consistent with the star-formation episodes.}

\keywords{Galaxies: elliptical and lenticular, cD; Galaxies: individual: NGC\,3607;
Galaxies: fundamental parameters}

\titlerunning{Stellar population, metallicity and ionised gas in NGC\,3607}

\authorrunning{M.G. Rickes et al.} 

\maketitle

\section{Introduction}
\label{Intro}

Significant efforts have been undertaken in the last few years to understand the
complex mechanisms associated with the formation and evolution of the early-type 
galaxies. Since the early 90's it was established that ellipticals cannot be described as 
being part of a one-parameter family, and that many show unambiguous signatures of 
interaction with companions and the environment (\citealt{vdB90}). Such facts suggest 
that these galaxies may have experienced different star formation histories, with 
the present-day stellar populations differing in metallicity and/or age. In this sense, 
most early-type galaxies (or their nuclei) are not characterised by a single-aged stellar 
population. Instead, they are better described by a composition of single stellar 
populations (e.g. \citealt{wor92}; \citealt{arim87}; \citealt{rickes2004}; 
\citealt{rickes2008}).

Observationally, clues to the star formation history of a galaxy can be found in the 
radial dependence of the stellar $\rm[\alpha/Fe]$ ratio, or more specifically the Mg/Fe 
abundance ratio (\citealt{pipino2006}). The $Mg2$ line-strength distribution in early-type 
galaxies can vary considerably, with radial profiles ranging from essentially featureless 
to rather structured. In the latter case, changes of slope, possibly associated with 
kinematically decoupled cores, or anomalies in the stellar population, have been observed 
in some ellipticals  (e.g. \citealt{car93}).

In this context, the investigation of the spatial distribution of the stellar population 
and metallicity in samples of early-type galaxies may shed some light on the formation and 
evolution issues. With the above goals in mind, we have been conducting a systematic study 
of objects rich in interstellar medium, such as NGC\,5044 (\citealt{rickes2004}) and NGC\,6868
(\citealt{rickes2008}).

\begin{figure}
\resizebox{\hsize}{!}{\includegraphics{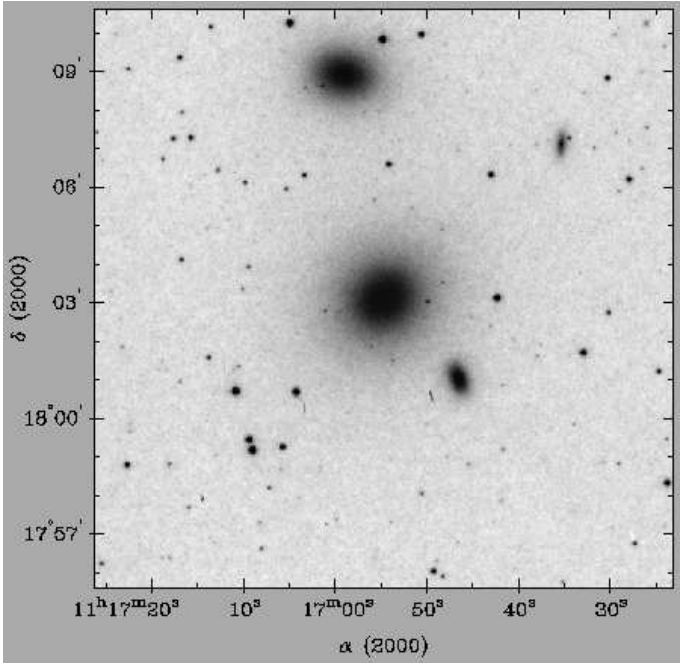}}
\caption{V band image covering the field $15\arcmin\times15\arcmin$ centred on NGC\,3607.
Image orientation: North to the top and East to the left.}
\label{fig1}
\end{figure}

This work focus on the lenticular galaxy NGC\,3607, which is the central and brightest member 
of a group that contains the galaxies NGC\,3605 and NGC\,3608 as well. Fig.~\ref{fig1} shows a 
$15\arcmin\times15\arcmin$ V band image of NGC\,3607 and companions, extracted from the Canadian 
Astronomy Data Centre (CADC)\footnote{\em http://cadcwww.dao.nrc.ca/}.

The radial velocity of NGC\,3607 is $\sim960$ km\,s$^{-1}$ (\citealt{for90}), which puts it 
at a distance of $\sim12.8$\,Mpc (with $\rm H_0=75\,km\,s^{-1}~Mpc^{-1}$). At such a distance, 
1\arcsec\ corresponds to about 0.062\,kpc. NGC\,3607 is a LINER with X-ray emission
(\citealt{terashima2002}), and discrete sources in the host galaxy are the probable origin of 
the extended hard X-ray emission. The hot gas in this elliptical galaxy has been analysed by
\citet{matsu2000}. NGC\,3607 is another case where the dust and ionised gas distribute as a 
small asymmetric disk. The dust absorption is stronger in the northeast direction, while the 
ionised gas emission is more important towards the southwest (\citealt{ferr99}).

The goal of this paper is to investigate the stellar population, metallicity distribution 
and the presence of ionised gas in NGC\,3607, which are important parameters to understand 
the formation and evolution of this galaxy.

This paper is organised as follows: Sect.~\ref{Observ} presents the observations and data reduction; 
Sect.~\ref{LickInd} deals with the equivalent width analysis of the absorption lines and their radial 
dependence; Sect.~4 describes the stellar population synthesis, whose results are discussed 
in Sect.~5; Sect.~6  presents an analysis of the emission lines and the nature of the ionisation 
source. Finally, in Sect.~7 a general discussion of the results and conclusions will be drawn.

\section{Observations and spectra extraction}
\label{Observ}

In the present paper we study the star formation history, metallicity distribution, and emission-gas 
properties of NGC\,3607. Basically, we work with long-slit spectroscopy and a star formation
synthesis algorithm, following similar methods as those applied to NGC\,6868 and NGC\,5903
(\citealt{rickes2008}). General parameters of the bright S0 galaxy NGC\,3607 are given in 
Table~\ref{tab1}. 

\begin{table}
\caption[]{General data on NGC\,3607}
\label{tab1}
\renewcommand{\tabcolsep}{4.0mm}
\renewcommand{\arraystretch}{1.25}
\begin{tabular}{lcl}
\hline\hline
Parameter&& Value\\
\hline
$\rm\alpha(J2000)$               && $11^h16^m54.6^s$ \\
$\rm\delta(J2000)$               && $+18\degr03\arcmin07\arcsec$ \\
$M_B$                            && $-21.92$                           \\
$B$                              && $10.8$                             \\
$E(B-V)$                         && $0.070$                            \\
$L_{H\alpha + [NII]}$            && $\rm1.28 \times 10^{41}\,erg\,s^{-1}$ \\ 
Heliocentric radial velocity     && $\rm960\pm20~km~s^{-1}$        \\
Distance                         && $12.8$\,Mpc\\
Redshift                         && $0.003202\pm0.000067$ \\
Diameters                        && $4.9\arcmin\times2.5\arcmin$        \\
Effective radius$^\dagger$       && $43.4\arcsec$\\
%$H_0$ ($\rm km\,s^{-1}~Mpc^{-1}$)  && $75$                  \\
\hline
\end{tabular}
\begin{list}{Table Notes.}
\item Data obtained at the NASA/IPAC Extragalactic Database (NED) which is operated by the Jet
Propulsion Laboratory, California Institute of Technology, under contract with the National 
Aeronautics and Space Administration. ($\dagger$) $R_e$ from \citet{Annibali07}.
\end{list}
\end{table}

The observations were  obtained 
with the ESO 3.6m telescope at La Silla, Chile, equipped with EFOSC1. The spectra cover 
the range $\lambda\lambda5100 - 6800$\,\AA\ with a 3.6\AA/pixel resolution. The spatial 
scale of the observational configuration is $0.6\arcsec\rm\,pixel^{-1}$; the slit length 
was $3.1\arcmin$ while its width was fixed at $1.5\arcsec$, corresponding approximately to 
the average seeing; two 2-dimensional spectra were obtained with the same exposition time 
(approximately 45 minutes) and  position angle $PA = 301\degr$. In all cases, the extraction
area was fixed at $4.57\arcsec^2$, which corresponds to $\rm\approx0.02\,kpc^2$.

The standard star $HR\,7333$ was observed for flux and velocity calibrations.
We point out that the average FWHM of the absorption features in the spectrum of 
$HR\,7333$ is between 8 and 10\,\AA, similar to that of the Lick/IDS system. 

\begin{figure}
\resizebox{\hsize}{!}{\includegraphics{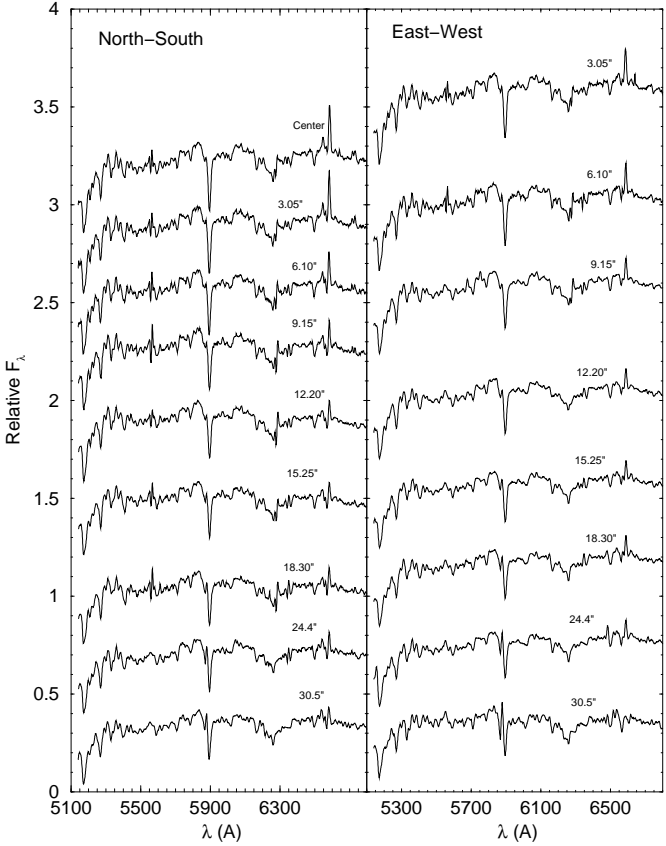}}
\caption{Spatial extractions for the SW (left panel) and NE (right panel) directions 
normalised at $\lambda5870$\,\AA. The distance to the galaxy centre is indicated on each 
spectrum. Except for the bottom ones, the spectra have been shifted by an arbitrary constant, 
for clarity purposes}
\label{fig2}
\end{figure}

The spectra were corrected for the radial velocity $\rm V_r = 960\,km\,s^{-1}$ and the 
foreground Galactic extinction $A_V = 0.035$ (Table~\ref{tab1}). The corrected spectra, 
normalised at $\lambda5870$\AA, are shown in Fig.~\ref{fig2}, separately for the northeast 
and southwest directions on the galaxy.
%                                                One column figure
%----------------------------------------------------------- S_vib

\section{Radial properties of absorption features}
\label{LickInd}

Important information on the chemical abundance and stellar population of a galaxy can be
obtained from the analysis of the equivalent widths of a set of selected absorption lines 
and continuum fluxes (e.g. \citealt{rickes2008}; \citealt{rickes2004}, and references 
therein). Here we work with the continuum tracing and spectral windows for the absorption
features defined by the Lick system (\citealt{fab85}). 

Because of the somewhat restricted coverage of our NGC\,3607 spectra (Sect.~\ref{Observ}), 
the available absorption features are \ion{Mg}{2}, \ion{Fe}{i}$_{\lambda5270}$, 
\ion{Fe}{i}$_{\lambda5335}$, \ion{Fe}{i}$_{\lambda5406}$, \ion{Fe}{i}$_{\lambda5709}$, 
\ion{Fe}{i}$_{\lambda5782}$, \ion{Na}{i}$_{\lambda5895}$, and \ion{TiO}{ii}$_{\lambda6237}$. 
Continuum points within the same spectral region are $\lambda_{5300}$, $\lambda_{5546}$, 
$\lambda_{5650}$, $\lambda_{5800}$, $\lambda_{5870}$, $\lambda_{6173}$, $\lambda_{6620}$, and 
$\lambda_{6640}$.

Some early-type galaxies have been shown to present systematic spatial variations on the 
line strength of the indices \ion{Mg}{2} and \ion{Fe}{i}, either from the centre to the 
external regions or limited to the central region (e.g. \citealt{gonz93}; \citealt{car94}; 
\citealt{car93}; \citealt{davis93}; \citealt{fish95}; \citealt{fish96}).

The above arguments suggest the occurrence of changes in some fundamental properties of the
underlying stellar population. In order to investigate the stellar population and metallicity 
gradient in NGC\,3607, we first analyse the spatial distribution of metal-line strengths, and 
then apply a stellar population analysis.

Strong absorption lines of neutral iron and sodium, as well as Mg2 and TiO bands, are present
in the spectra of NGC\,3607 (Fig.~\ref{fig2}). Particularly conspicuous features are the
\ion{Mg}{2}$_{\lambda5176}$,
\ion{Fe}{i}$_{\lambda5270}$, \ion{Fe}{i}$_{\lambda5335}$, \ion{Fe}{i}$_{\lambda5406}$, 
\ion{Fe}{i}$_{\lambda5709}$, \ion{Fe}{i}$_{\lambda5782}$, \ion{Na}{i}$_{\lambda5895}$, and 
\ion{TiO}{ii}$_{\lambda6237}$. These absorption features present a strong dependence with the age 
and metallicity of the underlying stellar population (e.g \citealt{bi86}; \citealt{rickes2004}; 
\citealt{rickes2008}).

Each equivalent width (EW) has been measured three times, allowing for the uncertainties 
in the continuum level definition due to noise in the spectra. This procedure allowed us to 
estimate the average value and corresponding standard deviation from the average for each 
measurement. 

Subsequently, the observed values of EW were corrected for line broadening due to the stellar 
velocity dispersion. We followed the same procedure as described in \citet{rickes2008}. The 
spectrum of the G giant star $HR\,7333$, assumed to have zero intrinsic velocity, was broadened 
with a series of Gaussian filters with velocity dispersion $\sigma$ varying from $0$ to $\rm
300\,km\,s^{-1}$ in steps of $\rm 10\,km\,s^{-1}$. For each absorption feature considered we 
calculate an empirical correction index ${C_\sigma}$, so that $C_\sigma(EW) =\frac{EW(0)}{EW(\sigma)}$,
where EW(0) and EW($\sigma$) are the observed and broadened equivalent widths, respectively. 
We calculate the correction index for each extracted spectrum using the $\sigma$ values defined 
by \citet{ca00}. 

The EWs measured in the spectra of NGC\,3607, corrected for the velocity dispersion, are given 
in Table~\ref{tab3}. Except for the Mg2 index, which is measured in magnitude, the remaining EWs 
are given in \AA. 

\begin{table*}
\caption[]{Equivalent widths measured in the spectra of NGC\,3607}
\label{tab3}
\renewcommand{\tabcolsep}{1.3mm}
\renewcommand{\arraystretch}{1.25}
\begin{tabular} {lcccccccccc}
\hline 
$R$& $R/R_e$ & $Mg_2$index & $Mg_{2\,\lambda5176}$ & $FeI_{\lambda5270} $ & $FeI_{\lambda5335}$ & 
$FeI_{\lambda5406}$ & $FeI_{\lambda5709}$ & $FeI_{\lambda5782}$ & $NaI_{\lambda5895}$ & 
$TiO_{II\,\lambda6237}$ \\
(\arcsec) & & (mag) &(\AA)&(\AA)&(\AA)&(\AA)&(\AA)&(\AA)&(\AA)&(\AA) \\
\hline    
0.00    & 0.00 & $0.17\pm0.01$ & $6.10\pm0.22$ & $3.60\pm0.17$ & $3.75\pm0.15$ & $2.27\pm0.04$ & $1.33\pm0.02$ & $1.01\pm0.05$ & $5.03\pm0.21$ & $6.61\pm0.33$ \\
3.05S   & 0.07 & $0.16\pm0.01$ & $6.03\pm0.25$ & $3.58\pm0.35$ & $3.29\pm0.26$ & $2.35\pm0.09$ & $1.34\pm0.04$ & $1.06\pm0.01$ & $5.36\pm0.22$ & $6.42\pm0.32$ \\
6.10S   & 0.14 & $0.16\pm0.01$ & $5.78\pm0.35$ & $3.46\pm0.34$ & $3.13\pm0.52$ & $2.38\pm0.03$ & $1.43\pm0.06$ & $1.15\pm0.01$ & $4.76\pm0.25$ & $6.41\pm0.35$ \\
9.15S   & 0.21 & $0.15\pm0.01$ & $5.49\pm0.38$ & $3.45\pm0.28$ & $3.38\pm0.20$ & $2.08\pm0.11$ & $1.30\pm0.03$ & $1.29\pm0.03$ & $4.72\pm0.30$ & $6.54\pm0.33$ \\
12.20S  & 0.28 & $0.15\pm0.01$ & $5.56\pm0.38$ & $3.08\pm0.24$ & $3.07\pm0.29$ & $2.02\pm0.15$ & $1.02\pm0.10$ & $1.26\pm0.08$ & $4.04\pm0.27$ & $5.64\pm0.35$ \\
15.25S  & 0.35 & $0.14\pm0.01$ & $4.97\pm0.26$ & $2.80\pm0.48$ & $2.76\pm0.16$ & $2.35\pm0.06$ & $1.08\pm0.08$ & $1.12\pm0.02$ & $4.00\pm0.24$ & $5.65\pm0.38$ \\
18.30S  & 0.42 & $0.14\pm0.01$ & $4.95\pm0.40$ & $2.80\pm0.56$ & $2.73\pm0.30$ & $2.27\pm0.07$ & $1.06\pm0.11$ & $1.01\pm0.07$ & $3.99\pm0.31$ & $4.85\pm0.34$ \\
24.40S  & 0.56 & $0.13\pm0.01$ & $4.89\pm0.53$ & $3.14\pm0.49$ & $2.76\pm0.27$ & $1.86\pm0.14$ & $1.05\pm0.11$ & $1.33\pm0.13$ & $3.93\pm0.32$ & $5.63\pm0.37$ \\
30.50S  & 0.70 & $0.14\pm0.01$ & $4.92\pm0.62$ & $3.21\pm0.63$ & $2.65\pm0.57$ & $1.79\pm0.18$ & $1.23\pm0.12$ & $0.92\pm0.02$ & $3.43\pm0.32$ & $5.35\pm0.38$ \\
\hline      
3.05N   & 0.07 & $0.16\pm0.01$ & $6.03\pm0.20$ & $3.64\pm0.33$ & $3.43\pm0.17$ & $2.28\pm0.04$ & $1.31\pm0.04$ & $1.00\pm0.02$ & $5.05\pm0.23$ & $6.38\pm0.33$ \\
6.10N   & 0.14 & $0.16\pm0.01$ & $5.89\pm0.28$ & $3.54\pm0.37$ & $3.34\pm0.26$ & $2.26\pm0.15$ & $1.22\pm0.05$ & $1.03\pm0.03$ & $4.67\pm0.30$ & $6.36\pm0.35$ \\
9.15N   & 0.21 & $0.15\pm0.01$ & $5.52\pm0.08$ & $3.45\pm0.24$ & $3.15\pm0.32$ & $2.21\pm0.16$ & $1.30\pm0.04$ & $1.19\pm0.02$ & $4.96\pm0.27$ & $6.40\pm0.31$ \\
12.20N  & 0.28 & $0.14\pm0.01$ & $5.39\pm0.28$ & $3.18\pm0.27$ & $3.00\pm0.24$ & $2.06\pm0.09$ & $1.07\pm0.07$ & $1.18\pm0.02$ & $4.56\pm0.31$ & $5.55\pm0.33$ \\
15.25N  & 0.35 & $0.14\pm0.01$ & $5.09\pm0.16$ & $2.84\pm0.51$ & $2.55\pm0.24$ & $2.22\pm0.10$ & $1.11\pm0.12$ & $1.03\pm0.05$ & $4.08\pm0.25$ & $5.46\pm0.38$ \\
18.30N  & 0.42 & $0.15\pm0.01$ & $5.00\pm0.30$ & $2.81\pm0.45$ & $2.74\pm0.13$ & $2.21\pm0.10$ & $1.03\pm0.14$ & $0.95\pm0.02$ & $4.02\pm0.27$ & $4.78\pm0.35$ \\
24.40N  & 0.56 & $0.14\pm0.01$ & $4.93\pm0.23$ & $2.93\pm0.50$ & $2.74\pm0.26$ & $1.93\pm0.13$ & $1.02\pm0.05$ & $1.29\pm0.05$ & $3.89\pm0.30$ & $5.65\pm0.37$ \\
30.50N  & 0.70 & $0.13\pm0.01$ & $4.89\pm0.12$ & $3.10\pm0.60$ & $2.59\pm0.46$ & $1.82\pm0.23$ & $1.19\pm0.10$ & $0.98\pm0.01$ & $3.55\pm0.31$ & $5.47\pm0.35$ \\
\hline
\end{tabular}
\begin{list} {Table Notes.}
\item  Note that EW of Mg$_2$ index is given in mag. Extraction area of all spectra of NGC\,3607
is $4.57(\arcsec)^2\approx0.02\,kpc^2$. The EWs are corrected for velocity dispersion and  the errors are 
the standard deviation of the average of the three different EW measurements for each line.
\end{list}
\end{table*}

\begin{figure}
\resizebox{\hsize}{!}{\includegraphics{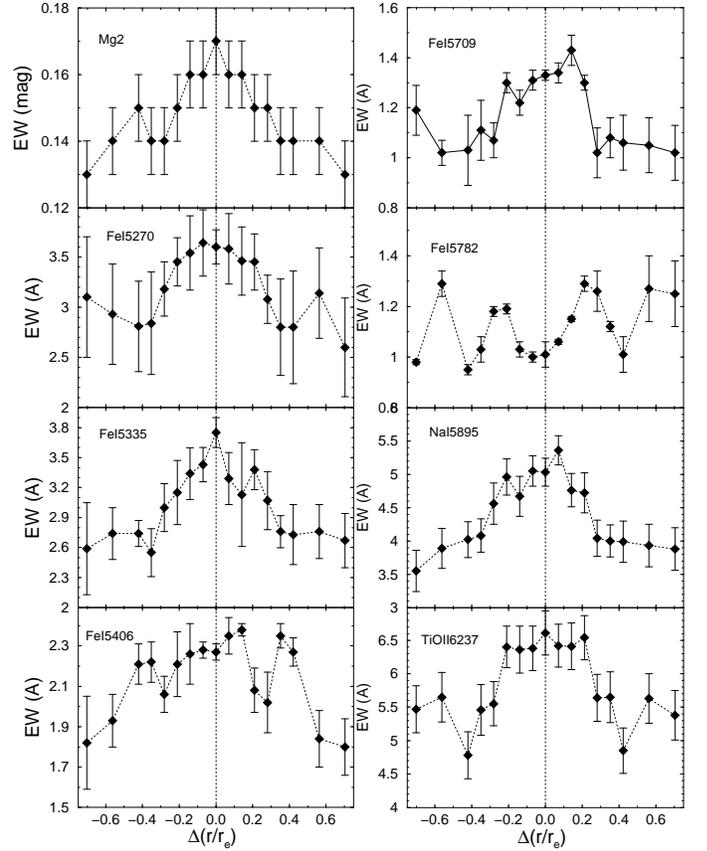}}
\caption{Radial distribution of the absorption features in NGC\,3607. The radial scale is
given in terms of the effective radius, $R_e=43.4\arcsec$, with negative values towards the
North direction.}
\label{fig3}
\end{figure}

The corrected EWs of NGC\,3607 (Table~\ref{tab3}) are shown as a function of the distance to the galaxy 
centre in Fig.~\ref{fig3}. Except for \ion{Fe}{i}$_{\lambda5782}$, the remaining absorption features 
present significant variations with distance to the centre. In particular, the \ion{Mg}{2} band clearly 
shows a negative gradient, i.e. increasing towards the centre of the galaxy. Evidence of similar negative 
gradients are seen in the other $\ion{Fe}{i}$ features.

\begin{figure}
\resizebox{\hsize}{!}{\includegraphics[angle=0]{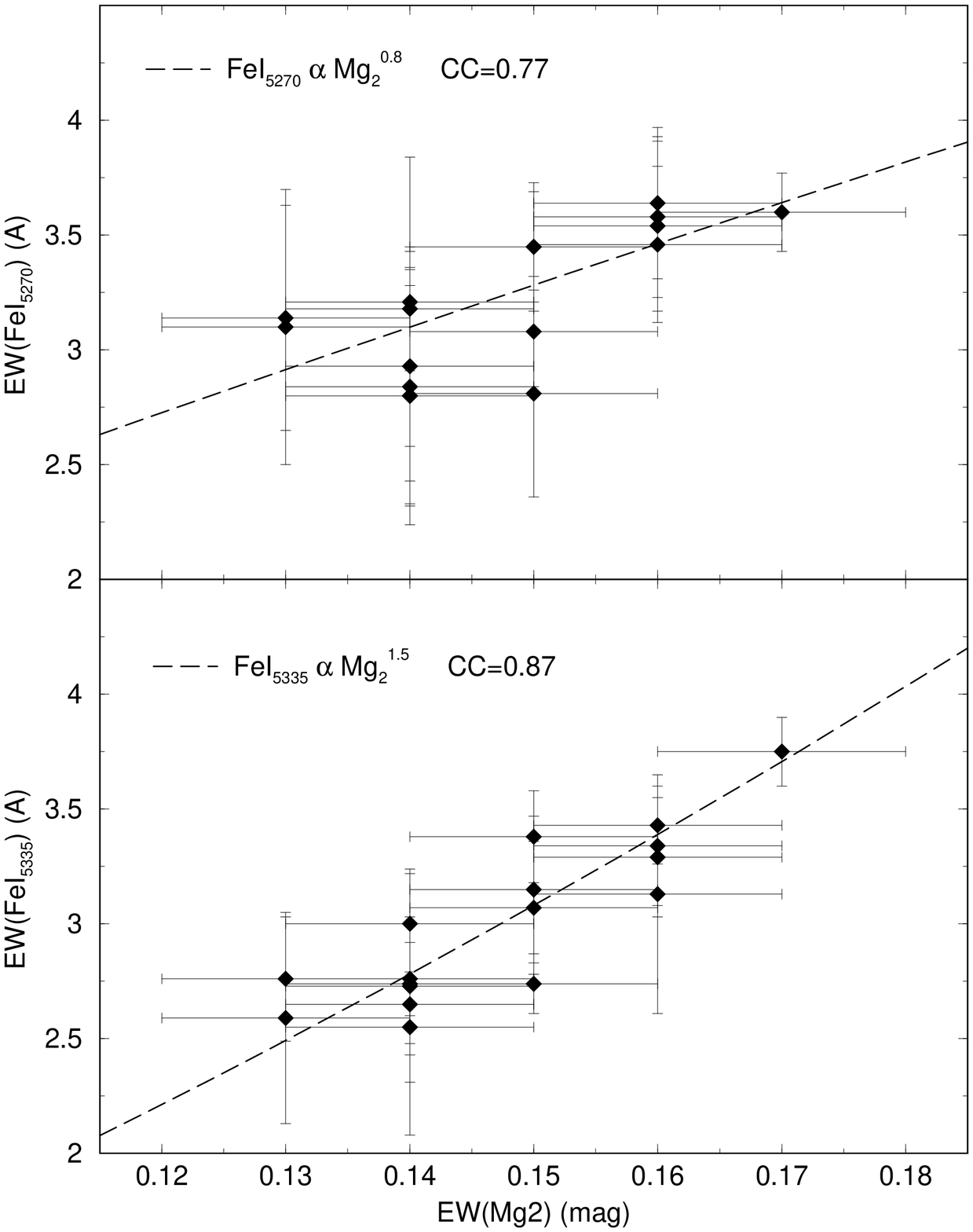}}
\caption{Correlations among the EWs of selected absorption features measured in NGC\,3607.
The correlation coefficients (CC) of the fits are indicated.}
\label{fig4}
\end{figure}

Similar gradient slopes in the \ion{Mg}{2} index and the \ion{Fe}{i} features suggest the 
same chemical enhancement process for these elements. Furthermore, as shown in Fig.~\ref{fig4}, 
the EW of \ion{Mg}{2} correlates both with $EW(\ion{Fe}{i}_{\lambda5270})$ and
$EW(\ion{Fe}{i}_{\lambda5335})$, which is consistent with the same enrichment process acting 
on Mg and Fe in NGC\,3607.

When the gravitational potential is strong enough to retain the gas ejected by supernovae
events and massive stars winds, it eventually migrates to the galaxy's central regions. Thus, 
it should be expected that subsequent generations of stars will be more metal rich at the 
centre than in the external parts. Consequently, a negative radial metallicity gradient should 
be established as a mass-dependent parameter. In this scenario, a galaxy formed essentially 
through a monolithic collapse, i.e. with a small contribution of mergers, is expected to present 
a well-defined correlation between the metallicity gradient(\citealt{lar74}; \citealt{ogan05})  
and total mass (\citealt{car93}).

The above issue was investigated by \citet{car93}, who studied the dependence of the gradient 
$d{\rm Mg_2}/d\log\,r$ with galaxy mass in a sample of 42 elliptical galaxies. They detected 
a conspicuous correlation for masses lower than $\sim10^{11}\,\ms$, followed by a saturation 
for higher masses. Different formation mechanisms, such as a less dissipative collapse and/or 
merger of smaller galaxies, are invoked to explain the lack of $Mg_2$ gradient slope to galaxy 
mass correlation for the high-galaxy mass range (e.g. \citealt{car93}). In addition, the 
correlation between $d{\rm Mg_2}/d\log\,r$ with velocity dispersion ($\sigma$) - confined to 
the low-mass galaxies found by \citet{ogan05} - suggest that only galaxies below some limiting 
$\sigma$ have been formed by collapse, whereas the massive ones were formed predominantly by 
mergers. 

In order to check how NGC\,3607 fits in the metallicity gradient {\em vs.} galaxy mass picture
of \citet{car93}, we first estimate the galaxy's dynamical mass with the approximation of
\citet{Poveda75}, $M_{tot} = 0.9\,\sigma^2\,R_e/G =3.34\times10^8\,\ms$, where $G$ is the 
gravitational constant, $R_e = 28.53\arcsec$ is the effective radius and 
$\rm\sigma = 225\,km\,s^{-1}$ is the central stellar velocity dispersion. The dynamical 
mass computed for NGC\,3607 is $M=(1.6\pm0.3)\times10^{11}\,\ms$ which, together with the 
\ion{Mg}{2} index gradient slope $dMg_2/d\log\,r=-0.02$, puts this galaxy in the merger-dominated
region (see Fig.~11 of \citealt{car93}).

\section{Stellar population synthesis}
\label{StellarPop}

In this section we investigate the star formation history and derive properties of the 
stellar population of NGC\,3607 by means of the stellar population synthesis method of 
\citet{bi88}. As population templates we use the synthetic star cluster spectra of
\citet{bruzual03}\footnote{Built with a \citet{salpeter55} Initial Mass Function}. 

As a first approach we tested large bases with templates with ages between 1\,Myr and 
13\,Gyr, but we found that the flux contribution of the components younger than a few 
$10^2$\,Myr is statistically meaningless. Thus, for simplicity we built a spectral base 
containing 7 templates. After several tests we verified that the base that best 
reproduces the observed spectra, without redundancy, is	the one with the ages 1, 5 and 
13\,Gyr, together with the metallicities $Z=0.008,\,\,0.02, 
\,\,{\rm and}\,\,0.05$.   Such values are taken as representative of the sub-solar, solar
and above-solar metallicity ranges (e.g. \citealt{rickes2008}). The distribution of 
templates in the age-metallicity plane is shown in Table~\ref{tab4}. We note that the 
working base is rather restricted in components to minimise potential age/metallicity 
degeneracies (e.g. \citealt{rickes2008}) due to the relatively small number of absorption 
features available in our spectra of NGC\,3607 (Fig.~\ref{fig2}).

We now compare the present stellar population synthesis results with previous works. 
Recentelly, \citet{Annibali07} have studied the underlying stellar population of a large 
sample of early-type galaxies, among them NGC\,3607, by means of the $H\beta$, Iron and Mgb 
age and metallicity indicators. They report for this galaxy an average stellar population age 
of 3.1\,Gyr, while we found an average age of 8.8\,Gyr, based on the the central and external 
region values, a value that is more consistent for an SO galaxy. This discrepancy could be due 
to the different methodologies used to obtain the age of NGC\,3607. Interestingly, we note
that this does not occur with other galaxies of \citet{Annibali07} that have been studied by 
us following the same techniques as in the present paper. For NGC\,5044 and NGC\,6868 
(\citealt{rickes2004, rickes2008}) we obtained the average ages of 13 and 9.8\,Gyr, respectively,
which are very close to the 14.2 and 9.2\,Gyr found by \citet{Annibali07}.

%=================================== tabela da base espectral ================================================

\begin{table}
\caption{Age and metallicity components}
\label{tab4}
\renewcommand{\tabcolsep}{0.52cm}
\renewcommand{\arraystretch}{1.25}
\begin{tabular} {rccc}
\hline\hline
      &$Z = 0.05$ & $Z = 0.02$ & $Z = 0.008$ \\
\hline
13\,Gyr & A1  & A2  & A3  \\
 5\,Gyr & B1  & B2  & B3  \\
 1\,Gyr &     & C2  &     \\
\hline
\end{tabular}
\end{table}

The stellar population synthesis algorithm provides directly the flux fractions (relative to 
the flux at $\rm\lambda 5870\,\AA$) that each age and metallicity template contributes to the 
observed spectrum. The sum of the template spectra, according to the individual flux fractions, 
should be representative of the stellar population contribution to the observed spectrum. A 
detailed description of the algorithm is provided in \citealt{rickes2004}.

\begin{figure}
\resizebox{\hsize}{!}{\includegraphics[angle=0]{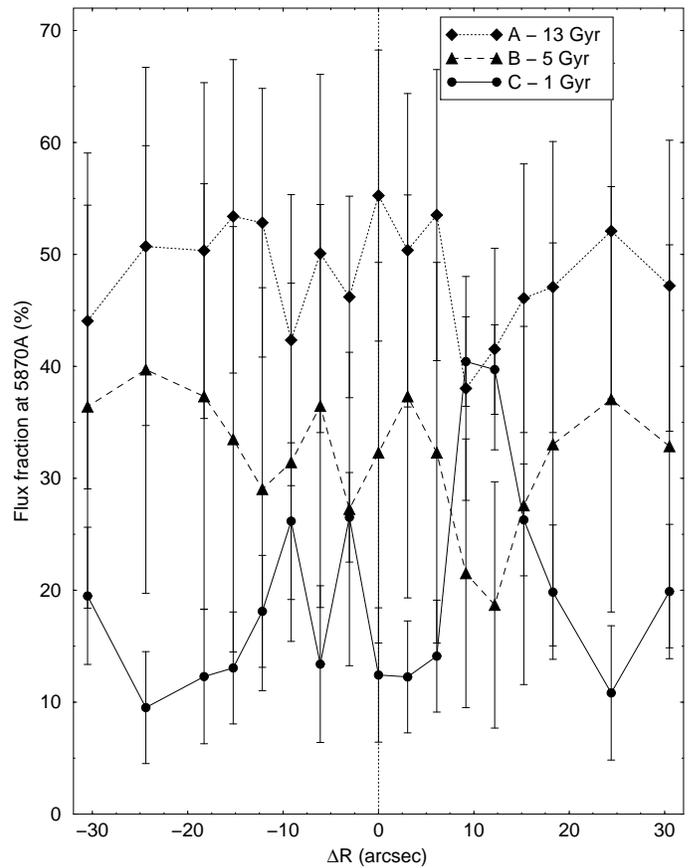}}
\caption{Flux-fraction contribution at $5870\,\AA$ of the stellar-population templates to the 
spectra of NGC\,3607. Negative values towards the North direction.}
\label{fig5}
\end{figure}

Results of the stellar population synthesis are summarised in Fig.~\ref{fig5}, which shows the 
1, 5, and 13\,Gyr component contributions to the flux at $5870\,\AA$, along most of the galaxy's 
radial extent. The 5 and 13\,Gyr curves in Fig.~\ref{fig5} correspond to the sum over the 
three metallicity components. In all cases the solar metallicity is the dominant component with 
a contribution higher than 70\%, while the $Z=0.05$ component contributes with $\approx10\%$, 
and the $Z=0.008$ with the remaining $\approx20\%$. Within uncertainties, the 5 and 13\,Gyr 
components dominate the spectra, from the centre to the external parts. The error bars in 
Fig.~\ref{fig5} are the standard deviation from the average of all possible solutions 
for each stellar population template.

As a caveat we note that our observed spectra of NGC\,3607 cover a relatively short spectral 
range. Thus, they contain a restricted number of indices sensitive to age and metallicity that, 
in principle, could provide a small number of constraints for the synthesis, especially in 
terms of metallicity. Nevertheless, Fig~\ref{fig5} indicates that the 13\,Gyr population 
dominates in NGC\,3607, contributing with approximately $50$\% of the flux at $5870\,\AA$. 
The 5\,Gyr population contributes with approximately $30$\%.

\begin{figure}
\resizebox{\hsize}{!}{\includegraphics[angle=0]{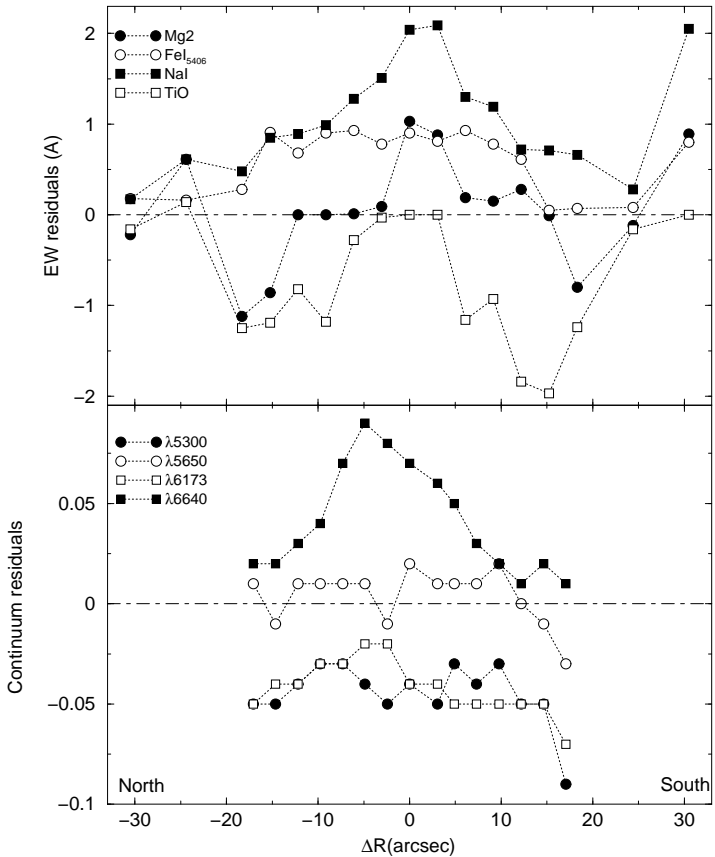}}
\caption{Stellar population synthesis residuals ($\rm EW_{\lambda}(obs)-EW_{\lambda}(synt)$) 
of representative EWs (top panel) and continuum points (bottom). Negative values towards the 
North direction.}
\label{fig6}
\end{figure}

Another way to assess the efficiency of the stellar population synthesis algorithm is by 
means of the residuals (difference between the observed and synthesised values), both in 
EWs and continuum points. With a few exceptions, the synthesised EWs are within $\pm1$\,\AA\
of the observed values, while for the synthesised continuum points (normalised to 
$\lambda5870$\,\AA) the offset lies within $\pm0.1$. We illustrate this in Fig.~\ref{fig6} 
for representative features and continuum points. The residuals are almost equally distributed 
between positive and negative values, which indicates that metallicity variations in NGC\,3607 
are essentially contemplated by the adopted templates (Table~\ref{tab4}).  

\section{Metallicity}
\label{metallicity}

We also investigate metallicity properties of NGC\,3607 by comparing the measured Lick indices 
of \ion{Mg}{2}, \ion{Fe}{i}$_{\lambda5270}$, and \ion{Fe}{i}$_{\lambda5335}$ (Table~\ref{tab3}), 
with the equivalent ones derived from single-aged stellar population (SSP) models 
(\citealt{thomas03}, \citealt{buzzoni94}) that assume a \citet{salpeter55} initial mass function
and different ages. After testing a range of SSP ages, we found that the indices measured in 
NGC\,3607 are best described by the 10\,Gyr models.

%========================================== modelos ssp ================================
\begin{figure}
\resizebox{\hsize}{!}{\includegraphics[angle=0]{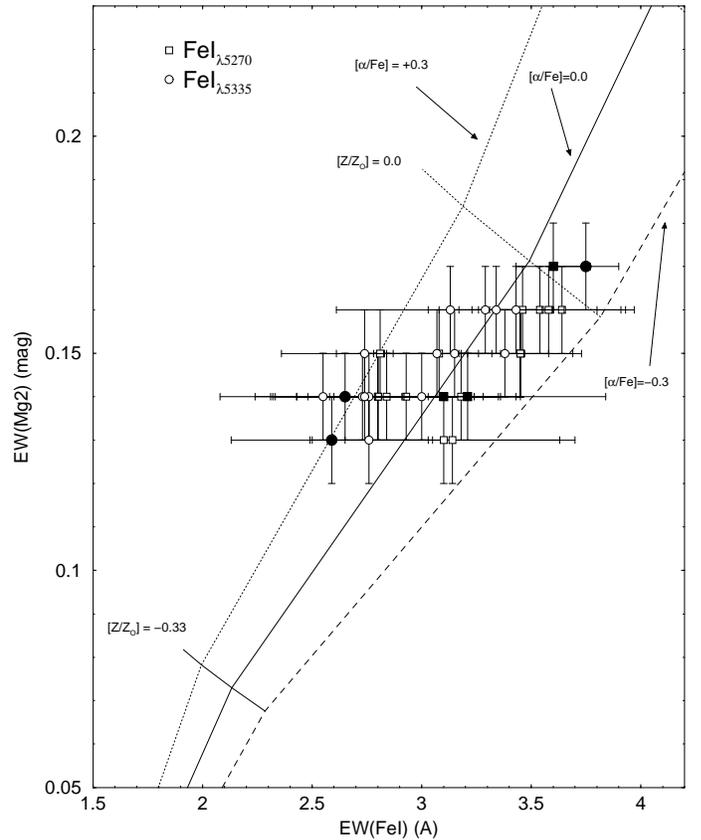}}
\caption{Selected Lick indices measured in NGC\,3607 are compared to those computed from SSP 
models for a range of metallicity and $\rm [\alpha/Fe]$ ratios. Filled symbols indicate the 
central (higher values) and external regions (lower values), respectively.}
\label{fig7}
\end{figure}

In Fig.~\ref{fig7} we compare our Lick indices measured in NGC\,3607 with the equivalent 
ones computed from SSP models built for different metallicities and $\rm [\alpha/Fe]$ ratios.
The central parts of NGC\,3607 present a small deficiency of alpha elements with respect to
iron ($\rm-0.3\la[\alpha/Fe]\la0.0$), whereas the opposite trend appears to characterise the 
external parts.  We found the average value of $\rm [\alpha/Fe]$ = 0.13 , which is comparable  
to the value 0.24 reported by \citet{Annibali07}.
	
NGC\,3607 presents sub-solar metallicity, with the external parts being more metal-poor 
than the nucleus. The differences observed in $\rm [\alpha/Fe]$ imply different mechanism of 
chemical enrichment for the central and external regions of NGC\,3607.

\section{Ionised gas}
\label{gas}

Emission gas has been detected in a large number of elliptical and lenticular galaxies 
\citep{phi86,macchetto96, rickes2004,rickes2008}. However, the origin of this gas and the 
ionisation source are not yet conclusively established. It is clear from Fig.~\ref{fig2} that
NGC\,3607 presents conspicuous emission lines throughout its body. In this section we investigate 
the physical conditions and ionisation source of the gas.

The properties of the emission gas are derived from line fluxes measured  in the stellar
population subtracted spectra (Sec~\ref{StellarPop}). The emission line fluxes of \ion{H}{$\alpha$}, 
\ion{[N}{ii]}$_{\lambda6548}$, \ion{[N}{ii]}$_{\lambda6584}$, \ion{[S}{ii]}$_{\lambda6717}$ and 
\ion{[S}{ii]}$_{\lambda6731}$ were measured with Gaussian functions fitted to the profiles. They 
are listed in Tab.~\ref{tab7}

\begin{table*}
\caption{Emission line parameters of NGC\,3607}
\label{tab7}
\renewcommand{\tabcolsep}{0.48cm}
\renewcommand{\arraystretch}{1.25}
\begin{tabular}{lccccccc}
\hline\hline
R (\arcsec) & $H\alpha$ & $[NII_{6584}]$ &  $[NII_{6548}]$ & $[SII_{6717}]$ & $[SII_{6731}]$ & 
$\frac{[NII]}{H\alpha}$  & $\frac{[SII]}{H\alpha}$ \\
(1)&(2)&(3)&(4)&(5)&(6)&(7)&(8)\\
\hline
0.00   & $1.34\pm0.1$ & $3.46\pm0.1$ & $1.15\pm0.1$ & $0.66\pm 0.1$ & $0.69\pm0.1$ & $3.44\pm0.1$ & $1.01\pm0.1$  \\
3.05S  & $1.29\pm0.1$ & $3.20\pm0.1$ & $1.06\pm0.1$ & $0.58\pm 0.2$ & $0.72\pm0.2$ & $3.30\pm0.1$ & $1.01\pm0.1$  \\
6.10S  & $0.93\pm0.1$ & $2.20\pm0.1$ & $0.73\pm0.1$ & $0.47\pm 0.2$ & $0.50\pm0.2$ & $3.15\pm0.1$ & $1.04\pm0.1$  \\
9.15S  & $0.60\pm0.2$ & $1.43\pm0.2$ & $0.47\pm0.2$ & $0.32\pm 0.2$ & $0.28\pm0.2$ & $3.17\pm0.1$ & $1.00\pm0.1$  \\
12.20S & $0.56\pm0.2$ & $1.10\pm0.2$ & $0.36\pm0.2$ & $0.14\pm 0.1$ & $0.32\pm0.1$ & $2.61\pm0.1$ & $0.82\pm0.1$  \\
15.25S & $0.60\pm0.2$ & $1.02\pm0.2$ & $0.34\pm0.2$ & $0.22\pm 0.2$ & $0.35\pm0.2$ & $2.27\pm0.1$ & $0.95\pm0.1$  \\
18.30S & $0.84\pm0.2$ & $1.25\pm0.2$ & $0.41\pm0.2$ & $0.19\pm 0.2$ & $0.36\pm0.2$ & $1.98\pm0.1$ & $0.65\pm0.1$  \\
24.40S & $0.80\pm0.2$ & $1.07\pm0.3$ & $0.35\pm0.2$ & $0.28\pm 0.2$ & $0.30\pm0.2$ & $1.78\pm0.1$ & $0.73\pm0.1$  \\
30.50S & $0.85\pm0.3$ & $0.89\pm0.3$ & $0.29\pm0.3$ & $0.26\pm 0.2$ & $0.31\pm0.2$ & $1.39\pm0.1$ & $0.67\pm0.1$  \\
\hline
3.05N  & $0.97\pm0.1$ & $2.43\pm0.1$ & $0.81\pm0.1$ & $0.42\pm 0.2$ & $0.44\pm0.2$ & $3.34\pm0.1$ & $0.89\pm0.1$  \\
6.10N  & $0.71\pm0.1$ & $1.87\pm0.1$ & $0.62\pm0.1$ & $0.41\pm 0.2$ & $0.47\pm0.2$ & $3.51\pm0.1$ & $1.24\pm0.1$  \\
9.15N  & $0.66\pm0.2$ & $1.43\pm0.2$ & $0.47\pm0.2$ & $0.31\pm 0.2$ & $0.31\pm0.2$ & $2.88\pm0.1$ & $0.94\pm0.1$  \\
12.20N & $0.65\pm0.2$ & $1.16\pm0.2$ & $0.38\pm0.2$ & $0.26\pm 0.2$ & $0.26\pm0.2$ & $2.37\pm0.1$ & $0.80\pm0.1$  \\
15.25N & $0.80\pm0.2$ & $1.07\pm0.3$ & $0.35\pm0.2$ & $0.30\pm 0.2$ & $0.30\pm0.2$ & $1.78\pm0.1$ & $0.75\pm0.1$  \\
18.30N & $0.90\pm0.3$ & $1.27\pm0.3$ & $0.42\pm0.3$ & $0.22\pm 0.3$ & $0.22\pm0.3$ & $1.88\pm0.1$ & $0.49\pm0.1$  \\
24.40N & $0.51\pm0.4$ & $1.03\pm0.4$ & $0.34\pm0.4$ & $0.42\pm 0.3$ & $0.42\pm0.3$ & $2.69\pm0.4$ & $1.65\pm0.2$  \\
\hline
\end{tabular}
\begin{list} {Table Notes.}
\item $H\alpha$ flux in col.~2 is given in $\rm 10^{-15}\,erg\,s^{-1}\,cm^{-2}$. Col.~7: 
$[NII]_{\lambda6584+6548}$; col.~8: $[SII]_{\lambda6717+6731}$. 
\end{list}
\end{table*}

\begin{figure}
\resizebox{\hsize}{!}{\includegraphics[angle=0]{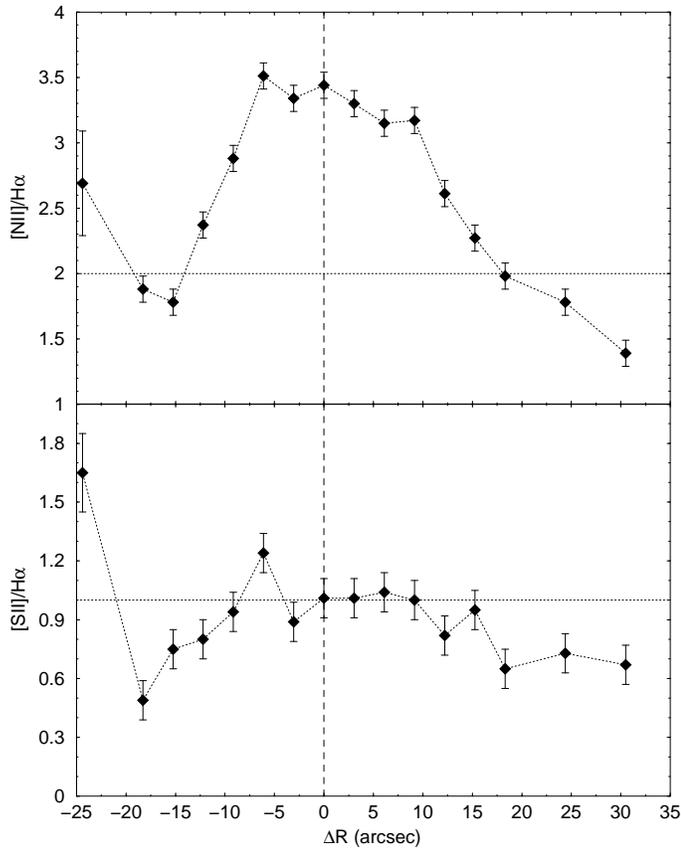}}
\caption{Spatial variation of the ratios $\frac{[NII]_{\lambda6584}}{H\alpha}$ (a) and $\frac{[SII]_{\lambda6731}}{H\alpha}$
(b).  Negative values towards the North direction.}
\label{fig8}
\end{figure}

The spatial distributions of the measured ratios $[NII]/H\alpha$ and $[SII]/H\alpha$ are shown in 
Fig.~\ref{fig8}. While the former presents significant changes with distance to the centre, the
latter is more uniformly distributed.

\begin{figure}
\resizebox{\hsize}{!}{\includegraphics{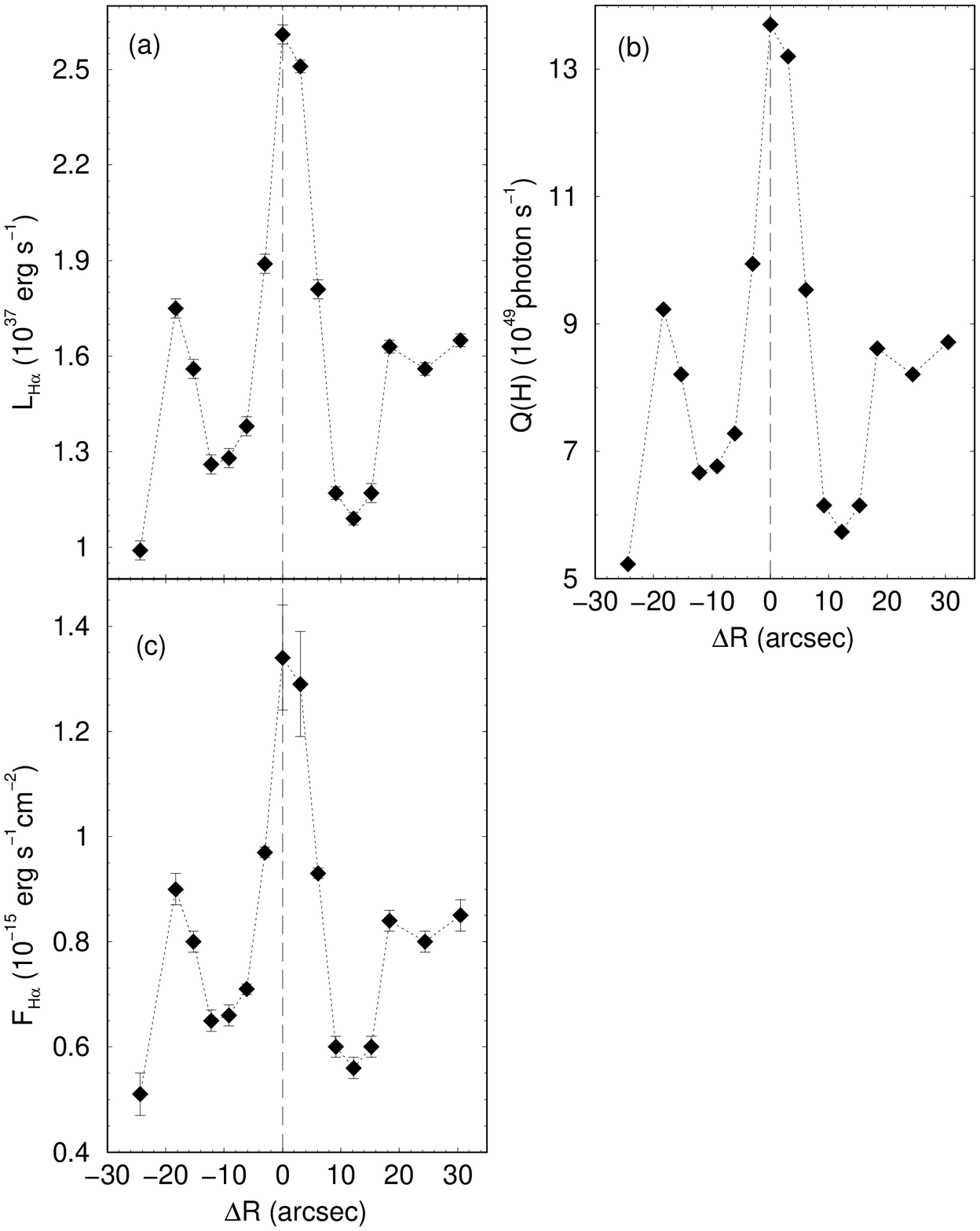}}
\caption{Spatial distribution of the $H\alpha$ luminosity (a) flux (c) and 
number of ionising photons (b). Negative values towards the North direction. }
\label{fig9}
\end{figure}

Assuming that the emission lines are formed essentially by recombination, the number of
ionising photons (Q(H)) can be  calculated as $Q(H) =\frac{L_{H\alpha}}{h\nu_\alpha}\frac{\alpha_B(H^0,T)}
{\alpha_{H\alpha} (H^0,T)}$ \citep{oster89},  where $\alpha_B\,(H^0,T)$ is the
total recombination coefficient, and $\alpha_{H\alpha}\,(H^0,T)$ is the recombination coefficient
for ${\rm H}\alpha$. 
The values of $Q(H)$ and $L_{H\alpha}$ are given in Tab.~\ref{tab8}, and their spatial distributions 
are shown in panels (a) and (b) of Fig.~\ref{fig9}. 

\begin{table}
\caption{$H\alpha$ luminosity and number of ionising photons}
\label{tab8}
\renewcommand{\tabcolsep}{0.23cm}
\renewcommand{\arraystretch}{1.25}
\begin{tabular}{lccc}
\hline\hline
 R   &  $F_{{H\alpha}}$ & $L_{H\alpha}$ & $Q(H)$ \\
(\arcsec) & $(\rm 10^{-15} erg\,cm^{-2}s^{-1})$ & $(\rm 10^{37} erg\,s^{-1})$ & $(\rm 10^{49}~s^{-1})$ \\
\hline
0.00S   & $1.34\pm0.10$ & $2.61\pm0.03$ & $13.7\pm0.02$ \\
3.05S   & $1.29\pm0.10$ & $2.51\pm0.02$ & $13.2\pm0.02$ \\
6.10S   & $0.93\pm0.01$ & $1.81\pm0.03$ & $9.54\pm0.01$ \\
9.15S   & $0.60\pm0.02$ & $1.17\pm0.02$ & $6.15\pm0.02$ \\
12.20S  & $0.56\pm0.02$ & $1.09\pm0.02$ & $5.74\pm0.03$ \\
15.25S  & $0.60\pm0.02$ & $1.17\pm0.03$ & $6.15\pm0.03$ \\
18.30S  & $0.84\pm0.02$ & $1.63\pm0.02$ & $8.62\pm0.03$ \\
24.40S  & $0.80\pm0.02$ & $1.56\pm0.02$ & $8.21\pm0.01$ \\
30.50S  & $0.85\pm0.03$ & $1.65\pm0.02$ & $8.72\pm0.02$ \\
\hline
3.05N   & $0.97\pm0.01$ & $1.89\pm0.03$ & $9.95\pm0.02$ \\
6.10N   & $0.71\pm0.01$ & $1.38\pm0.03$ & $7.28\pm0.01$ \\
9.15N   & $0.66\pm0.02$ & $1.28\pm0.03$ & $6.77\pm0.01$ \\
12.20N  & $0.65\pm0.02$ & $1.26\pm0.03$ & $6.67\pm0.02$ \\
15.25N  & $0.80\pm0.02$ & $1.56\pm0.03$ & $8.21\pm0.02$ \\
18.30N  & $0.90\pm0.03$ & $1.75\pm0.03$ & $9.23\pm0.02$ \\
24.40N  & $0.51\pm0.04$ & $0.99\pm0.03$ & $5.23\pm0.01$ \\
\hline
\end{tabular}
\begin{list} {Table Notes.}
\item $H\alpha$ luminosity and number of ionising photons computed for the
extractions given in col.~1.
\end{list}
\end{table}

%**********************************************************************************
%**********************************************************************************
%

The central region of NGC\,3607 presents $\ion{[N}{ii]}/H\alpha=3.44$ and
$\ion{[S}{ii]}/H\alpha>1$. We note that the ratio $\ion{[N}{ii]}/H\alpha$ is 
higher than 1.0  for all sampled regions (panel (a)).

Three hypotheses for the nature of the ionising source in the external regions of NGC\,3607 are 
tested: the gas is ionised by {\it (i)} an \ion{H}{ii} region, {\it (ii)} hot, post-AGB stars, or 
{\it (iii)} an AGN. They are discussed individually below.

\subsection{\ion{H}{ii} region}

We build an \ion{H}{ii} region with a mass distribution that follows the initial mass function
of \citet{salpeter55}, for stellar masses in the range $0.1 - 100\,\ms$. The model which best 
reproduces the number of ionising photons measured in the region at $3.05\arcsec$ is described 
in Tab.~\ref{tab9}. This table provides the distribution of stars for a representative set of 
spectral type and mass; in the last column we give the number of ionising photons 
($N_\ion{Ly}{\alpha}$) produced by each star (\citealt{Dottori81}). The total number of ionising
photons can be computed from $N(Ly)=\int_{10}^{30}N_{Ly}(m)\phi(m) dm$. The relation between the 
number of ionising photons with stellar mass can be approximated as
$N_{Ly}(m)\approx3.9\times10^{39}\left(\frac{m}{\ms}\right)^{5.9}\rm s^{-1}$ (\citealt{rickes2008}).
Given the above, the adopted model \ion{H}{ii} region produces 
$\rm\sim2.6\times 10^{50}\,photons\,s^{-1}$, which is comparable to the observed value.  
However, with more than 20 O stars, such an \ion{H}{ii} region would be very young, massive, 
and luminous enough to have already been detected in previous works. Besides, the stellar 
population synthesis (Sect.~\ref{StellarPop}) does not indicate significant, recent star
formation in the sampled regions of NGC\,3607.

\begin{table}
\caption{$\ion{H}{ii}$ region model parameters}
\label{tab9}
\renewcommand{\tabcolsep}{3.1mm}
\renewcommand{\arraystretch}{1.25}
\begin{tabular} {l c c c}
\hline\hline
Spectral type & M & Number & $N_{\ion{Ly}{\alpha}}$ \\
 &($M_\odot$)& (stars)& ($\rm 10^{48}~photon\,s^{-1}$) \\
\hline
$O5$ & $30$ & $20 $ & $2.00  $ \\
     & $22$ & $41 $ & $0.25  $ \\
$B0$ & $17$ & $75 $ & $0.082 $ \\
     & $15$ & $101$ & $0.043 $ \\
     & $13$ & $142$ & $0.015 $ \\
     & $10$ & $264$ & $0.0021$ \\
\hline
\end{tabular}
\begin{list} {Table Notes.}
\item Col.~2: representative stellar mass; Col.~3: number of stars; Col.~3: number of ionising
photons.
\end{list}
\end{table}

\subsection{Post-AGB stars}
\label{PAGB}

The number of ionising photons emitted by a typical Post-AGB star is $\rm\sim10^{47}\,s^{-1}$ 
\citep{bine94}. Thus, about $10^4$ such stars would be necessary to produce the amount of photons 
measured in each extracted region. Consequently, for an extraction area of $\rm\approx0.02\,kpc^2$
(Sect.~\ref{Observ}), the projected surface-density of post-AGB stars would be about 
$\rm5\times10^5\,kpc^{-2}$, a number prohibitively high (e.g. \citealt{bine94}). Besides, the 
spatial profiles of $[\ion{N}{ii}]$ and \ion{H}{$\alpha$} extracted along the slit (Fig.~\ref{fig8}), 
show a ratio $\ion{N[}{ii]}/\ion{H}{\alpha}>2$ for most regions, especially towards the central 
parts. However, the relatively high emission-line ratios measured in NGC\,3607, especially in the 
central region, are not consistent with ionisation models based on hot old stars in elliptical 
galaxies (e.g. \citealt{bine94}). The models of \citet{bine94} do not reproduce as well the 
observed large ratios $\ion{[N}{ii]}/\ion{H}{\alpha}$ and $\ion{[S}{ii]}_{6731}/\ion{H}{\alpha}$.
Instead, high values for these emission-line ratios are typical of AGNs.

On the other hand, the lower ratios of $[NII]_{\lambda6584}/H\alpha$ and $[SII]_{6731}/H\alpha$
(Fig.~\ref{fig8}) together with the lower values of Q(H) (Fig.~\ref{fig9}) measured for large 
radii may, in part, be accounted for by hot old stars.

\subsection{Active nucleus}

The arguments that follow from the above discussion suggest that NGC\,3607 hosts an AGN.
We test this hypothesis with the emission-line diagnostic-diagram involving the ratios 
$[NII]_{\lambda6584}/H\alpha$ and $[SII]_{\lambda6717,31}/H\alpha$. For comparison we take 
the equivalent values measured in a sample of nearby galaxies that host a low-luminosity 
($\rm L_{H\alpha}<2\times10^{39}\,erg\,s^{-1}$) active nucleus (\citealt{ho97}). To minimise 
the possibility of uncertain classification, we restrict the analysis to the galaxies in 
\citet{ho97} sample with unambiguous classification as LINER in spiral, LINER in elliptical, 
Seyfert or Starburst. The restricted sample contains 128 galaxies (see \citealt{rickes2008}).

\begin{figure}
\resizebox{\hsize}{!}{\includegraphics[angle=0]{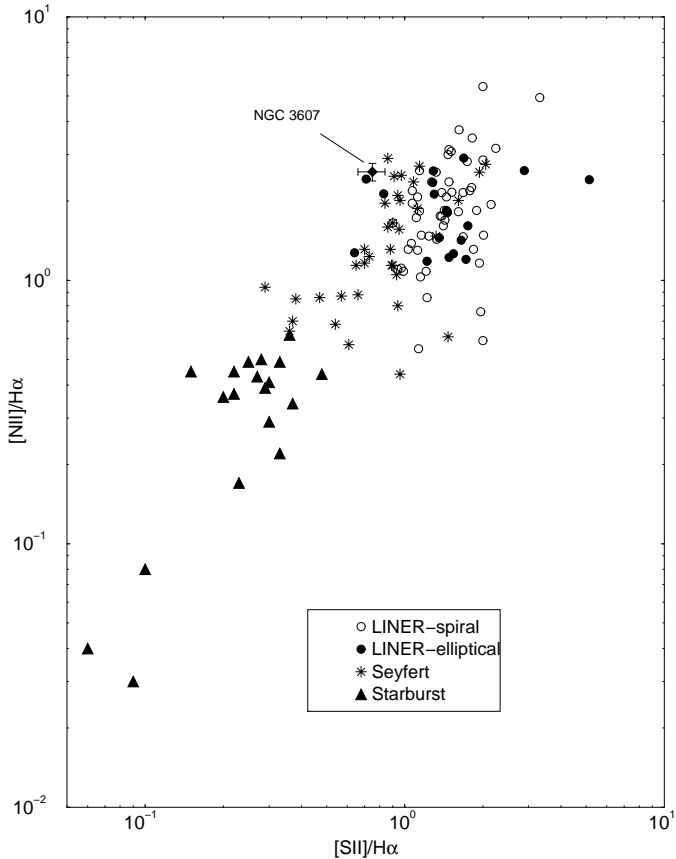}}
\caption{Emission-line diagnostic diagrams with a subsample of the galaxies in 
\citet{ho97} plotted as comparison. The locus of NGC\,3607 is consistent with that 
of LINERs.}
\label{fig10}
\end{figure}

The results are shown in Fig.~\ref{fig10}, where it can be seen that the values measured in 
NGC\,3607 occur in the locus occupied predominantly by LINERs and Seyferts. However, the lack 
of a broad component in $H\alpha$ (Fig.~\ref{fig2}) strongly favours the LINER interpretation. 
Together with the stellar population synthesis, this result suggests that the main source of 
gas ionisation in NGC\,3607 is non-thermal, produced by a low-luminosity active nucleus. The 
lower ratios of $[NII]_{\lambda6584}/H\alpha$ and $[SII]_{6731}/H\alpha$ towards the external 
parts can be accounted for by both a LINER and hot stars.

\begin{figure}
\resizebox{\hsize}{!}{\includegraphics[angle=0]{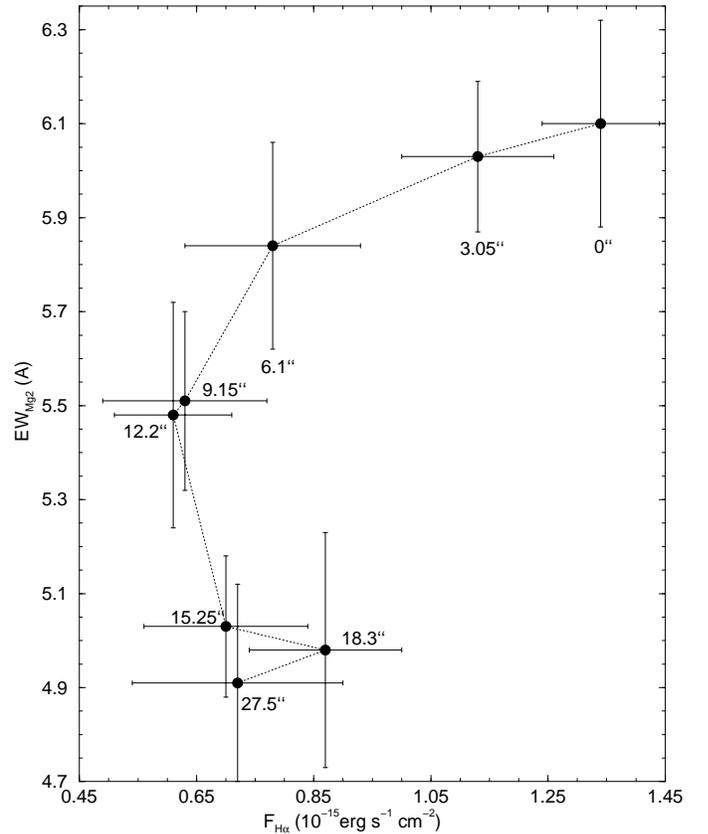}}
\caption{Relation between EW(Mg$_2~\lambda5176$) and the $H\alpha$ flux as a
function of distance to the galaxy centre.}
\label{fig11}
\end{figure}

Finally, in Fig.~\ref{fig11} we test whether the gradients in emission and 
absorption-line features are somehow related. We work with the EW of Mg$_2~\lambda5176$ 
(Table~\ref{tab3}) and the $H\alpha$ flux (Table~\ref{tab7}) distributions, taking
the average values of the North and South extractions (the 2 outermost points have
been averaged out to minimise error bars). Within the uncertainties, both distributions
appear to correlate for $R\la12\arcsec$, which can be accounted for the steep and
monotonic $H\alpha$ flux increase towards the central region (Fig.~\ref{fig9}) together 
with the overall negative Mg2 gradient (Fig.~\ref{fig3}). While the EW of Mg$_2$ keeps
dropping for $R\ga12\arcsec$, the $H\alpha$ flux oscillates abruptly, probably because 
of the relative contribution of the hot, old stars to the gas ionisation, superimposed 
on the severely diluted LINER flux at such large radii. This combination of effects is 
reflected on the mild anti-correlation of both properties (lower portion of Fig.~\ref{fig11}). 
It is also consistent with the low ratios of $\ion{[N}{ii]}/\ion{H}{\alpha}$ and
$\ion{[S}{ii]}/\ion{H}{\alpha}$ (Fig.~\ref{fig8}), and the low Q(H) (Fig.~\ref{fig9}),
measured for large radii, all accounted for by hot, old stars.

Similar results would be obtained with other absorption features, since they
also present similar gradients (Fig.~\ref{fig3}).

\section{Discussion and concluding remarks}
\label{Dicussion}

Individual galaxies can provide important pieces of evidence that may ultimately lead
to a better understanding of the formation and evolution processes of the early-type 
galaxies as a class. Following the methods previously employed in the analysis of 
other early-type galaxies (e.g. \citealt{rickes2004}; \citealt{rickes2004}), in the 
present paper we focus on the lenticular galaxy NGC\,3607.

We investigate the stellar population, metallicity distribution and ionised gas of
NGC\,3607 by means of long-slit spectroscopy (covering the wavelength range 
$\lambda\lambda5100-6800$\AA) and a previously tested stellar population synthesis 
algorithm. The stellar population synthesis shows different star formation episodes 
with ages distributed within 1-13\,Gyr.

The radial distribution of the equivalent widths measured for the available metallic 
absorption features in this galaxy consistently present a negative gradient, which 
indicates an overabundance of Fe, Mg, Na and TiO in the central parts with respect to 
the external regions. Also, the central region of NGC\,3607 presents a deficiency of 
alpha elements and higher metallicity than the external parts, which suggests different 
star-formation histories in both regions. With respect to the emission gas, the line 
ratios $[NII]_{\lambda6584}/H\alpha$ and $[SII]_{\lambda6717,31}/H\alpha$ consistently 
indicate the presence of a LINER at the centre of NGC\,3607.

We found that the $Mg_2$ index correlates both with ${\rm FeI}_{\lambda5270}$ and 
${\rm FeI}_{\lambda5335}$, which suggests that Mg and Fe probably underwent the same 
enrichment process in NGC\,3607. Also, the $H\alpha$ flux distribution correlates with 
the radial profile of metallic features, in the sense that they both decrease from the galaxy 
centre to $R\approx12\arcsec$. The correlation breaks for larger radii, especially because,
while the metallic feature profiles keep dropping, gas ionisation due to hot, old stars adds 
to the diluted LINER flux. Considering the relation between galaxy dynamical mass and the 
$Mg_2$ gradient slope (Figure\,11 of \citealt{car93}), the parameters computed for NGC\,3607 
in the present work suggest that this galaxy may have underwent important merger events in 
the past.

\section*{Acknowledgements}
We acknowledge important suggestions presented by the referee.
We thank the Brazilian agency CNPq for partial support of this work. This research has made 
use of the NASA/IPAC Extragalactic Database (NED) which is operated by the Jet Propulsion 
Laboratory, California Institute of Technology, under contract with the National Aeronautics 
and Space Administration.

%*********************************** REFERENCE ************************************

%**********************************************************************************
\end{document}